\documentclass{ws-procs9x6-cpt22}

\def\al{\alpha}
\def\be{\beta}

\def\de{\delta}

\def\et{\eta}\def\la{\lambda}

\def\ka{\kappa}
\def\rh{\rho}

\def\cG{{\cal G}}
\def\cL{{\cal L}}

\def\mn{{\mu\nu}}
\def\ab{{\al\be}}

\def\koo{{k_{00}}}
\def\kjj{{k_{jj}}}

\newcommand{\beq}{\begin{equation}}
\newcommand{\eeq}{\end{equation}}
\newcommand{\bea}{\begin{eqnarray}}
\newcommand{\eea}{\end{eqnarray}}
\newcommand{\rf}[1]{(\ref{#1})}
\def\etal {{\it et al.}}

\begin{document}


\title{New Signals in Precision Gravity Tests and Beyond}

\author{Q.G.\ Bailey$^1$, J.L.\ James$^2$, and J.R.\ Slone$^1$}

\address{$^1$Physics and Astronomy Department, Embry-Riddle Aeronautical University,
Prescott, AZ 86301, USA}
\address{$^2$Physics and Astronomy Department, Vanderbilt University,
Nashville, TN 37235, USA}

\begin{abstract}
We review the status of tests of
spacetime symmetries with gravity.  
Recent theoretical and experimental work
has involved gravitational wave signals,
precision solar-system tests, 
and sensitive laboratory tests searching
for violations of spacetime symmetries.
We present some new theoretical results
relevant for short-range gravity tests, 
with features of multiple length scales,
and possible large non-Newtonian forces 
at short distances.
\end{abstract}

\bodymatter

\section{Background}

Effective field theory has long been used to 
describe physics beyond General Relativity (GR) 
and the Standard Model (SM) of particle physics.\cite{weinberg09}
In particular, 
any speculated deviations from spacetime symmetry 
coming from an unknown fundamental theory\cite{ks89,review} 
can be described in a model-agnostic way 
by adding to GR and the SM, 
generic Lagrange density terms with
background coefficients and known field
operators.\cite{kp95}
This framework has been widely used and
can compare tests in distinctly different
regimes.\cite{sme1,k04,datatables}
The gravity sector of this effective field
framework has been studied for more than 16 years
and has revealed many intriguing signals
for new physics in tests such as gravimetry, 
lunar laser ranging, 
and gravitational waves.\cite{bk06,kt11,km16}

The gravity sector in the Riemann geometry limit 
is a general expansion with a Lagrange density 
in the form of
\beq
\cL = \frac {1}{\ka} \sqrt{-g} \left( R + s_\ab R^\ab + ... + k_\ab R^\ab R + ... \right),
\label{action1}
\eeq
where $g$ is the determinant of the metric $g_\mn$ 
and the first term with the Ricci scalar $R$ is GR, 
while the remaining terms form a series 
with coefficients for symmetry breaking 
(e.g., $s_\ab$) coupled to curvature tensors. 
The series may include dynamical terms for
the coefficients or the symmetry breaking may be explicit.\cite{bluhm,kl21,b21}

\section{Recent measurements and theory}

The phenomenology of the terms in
\rf{action1} has been studied in a number
of works.
Observable effects in weak-field gravity
tests have been established for minimal
and nonminimal terms,\cite{bk06,kt11,bkx15}
and some work has been done on strong-field 
gravity regimes like cosmology.\cite{bonder17,abn21,cos22}
Effects on gravitational waves have been
studied, 
showing that dispersion and birefringence
occur generically as a result of CPT and
Lorentz violation.\cite{km16}
Analysis has been performed in tests such as lunar laser ranging\cite{llr}, 
gravimetry\cite{gravi}, 
pulsars\cite{pulsars}, 
and using the catalog of GW events.\cite{gwtests}

On the theory side, 
explicit local Lorentz and diffeomorphism symmetry cases have been explored various contexts.
A ``3+1" or ADM formulation of the EFT framework 
has been explored
in Refs.\ \refcite{abn21,rs21}.
Also, 
explicit breaking solutions 
have been explored for phenomenology.\cite{bonder}
Other recent work includes much attention 
to vector models of spontaneous symmetry breaking, 
with black hole solutions being obtained\cite{bbmodels},
and the systematic construction of dynamical terms for the spontaneous symmetry breaking scenario
in the gravity sector.\cite{b21}
Finally we note some recent theoretical work 
has identified general properties of backgrounds 
in effective field theory,\cite{kl21} 
and new types of tests are possible that search for 
non-Riemann geometry.\cite{kl21nr}

\section{Short-range gravity signals}

Presently, 
the nature of gravity is unknown on length
scales less than micrometers.
New types of forces, stronger than the
Newtonian gravitational force,
could exist and be consistent with
experimental limits.\cite{srreview}

In references \refcite{bkx15} and \refcite{km17},
Lorentz-symmetry breaking solutions
for short-range gravity tests were found using
an approximation of first order in the 
coefficients for the modified Newtonian force.
This approximation makes searches 
in some short-range tests challenging, 
as some tests are designed 
to probe very small length scales 
at the cost of sensitivity to the Newtonian force.\cite{short}
We comment here on progress towards
``exact" solutions, i.e., to all orders in coefficients for Lorentz-symmetry breaking, 
that could be interesting for all short-range gravity tests.

One particular model that contains 
the interesting features of exact solutions
is the following Lagrange density, 
a subset of the general action in \rf{action1},
\beq
\cL = \frac {1}{\ka} \sqrt{-g} \left( R + k_{\al\be} R^{\al\be} R \right). 
\label{action2}
\eeq
The quantities $k_{\al\be}$ are the coefficients for Lorentz violation for this case. 

The action in \rf{action2}, 
taken in the linearized gravity limit, 
yields field equations for the metric fluctuations $h_\mn$.
If we further restrict attention to 
the static limit and only isotropic
coefficients $k_{00}$ and $k_{jj}$, 
in a special coordinate system,
we obtain the following coupled equations
for the metric components $h_{00}$ and $h_{jj}$:
\bea
\nabla^2 ( h_{00} + h_{jj} )
-3 (\koo- \frac {1}{9} \kjj ) \nabla^4 h_{00} 
+ (\koo - k_{kk} ) \nabla^4 h_{jj} &=& 
-32 \pi G_N \rh,
\nonumber\\
 \nabla^2 (3 h_{00} + h_{jj} ) 
+ 4 (\koo - \frac{1}{3} k_{kk} ) \nabla^4 h_{00} 
+ \frac {8}{3} k_{kk} \nabla^4 h_{jj}
&=& 0.
\label{coupled}
\eea
Note that $\kjj - \koo= k_\mn \et^\mn$ is a Lorentz invariant scalar combination. 

A Green function solution can be constructed
where we use a point source 
$4\pi G_N \rh = \de^{(3)}(\vec r - \vec r^\prime )$,
and the point source solutions 
for $h_{00}$ and $h_{jj}$ are denoted $\cG_1$ and $\cG_2$.
With $\vec R=\vec r - \vec r^\prime$,
and with guidance from 
standard catalogues of Green functions\cite{Lindell},
the general solutions take the form:
\bea
    \cG_1 = {1 \over R } 
    \left( A_1 e^{-q_{1} R}+A_{2} e^{-q_{2} R}+A_{3}\right),
    \nonumber\\
    \cG_2 = {1 \over R } \left( B_ 1 e^{-q_{1} R}+B_{2} e^{-q_{2} R}+B_{3}\right).
    \label{newpot}
\eea
Here the $A_n$'s and $B_n$'s are constants
to be solved for as well as the $q_1$ and $q_2$.  
In constructing this solution we are
assuming boundary conditions 
where the metric components 
vanish far from the source.
Insertion of \rf{newpot} into the 
point source version of \rf{coupled}, 
followed by using the well-known 
properties of the functions of 
$R=|\vec r - \vec r^\prime |$, 
allows one to solve for the $8$ parameters $A_1$, $A_2$, $A_3$, $B_1$, $B_2$, $B_3$,
$q_1$, and $q_2$ from $8$ resulting algebraic equations.

First, 
we find that for nontrivial solutions, 
both $q_1^2$ and $q_2^2$ must satisfy 
a quartic equation.
To display this, 
it is convenient to write a short-hand $q^2=u \pm v$, 
where $u$ and $v$ are given by:
\bea
u &=& { - (\koo - \frac 53 \kjj) \over 2 
(\koo + \frac 13 \kjj )^2 },
\nonumber\\
v &=& { \sqrt{(\koo - \frac 53 \kjj )^2 - 4 (\koo + \frac 13 \kjj )^2} \over 
2 (\koo + \frac 13 \kjj )^2 }
\label{uv}
\eea
Note that $u$ is real and $v$ can be complex.
The positions of the roots $\{ q=z^{1/2}=
\pm (u \pm v)^{1/2} \}$ in the complex
plane depend on the values of the
coefficients $\koo$ and $\kjj$.
If $q$ is entirely real and positive, 
then the solutions in \rf{newpot} will
exhibit exponential damping in $R$ 
or short-range Yukawa-like behavior.
The case where $q$ is negative and real
will result in runaway exponential increase 
and is not physically viable.
When $q$ has an imaginary piece 
or is entirely imaginary, 
the solution will have oscillations in $R$.

As a sample for this proceedings, 
we assume the condition $q_1^2 \neq q_2^2$.
This condition ensures that the coefficients
$\koo$ and $\kjj$
are treated {\it a priori} independent.
In \rf{uv}, 
$q_1^2$ then takes one sign in the $\pm$, 
and $q_2^2$ takes the other sign.
For this case we obtain the solutions for the Green function $\cG_1$ as follows:
\bea
\cG_1 &=& {1 \over 2\pi R}  
- {1 \over 4 \pi } \left( 1 + { \koo + \frac {11}{3} \kjj \over 
\sqrt{(\koo - \frac 53 \kjj )^2 - 4 (\koo + \frac 13 \kjj )^2} } \right)
{ e^{-R/\la_+} \over R} 
\nonumber\\
&&
- {1 \over 4 \pi } \left( 1 - { \koo + \frac {11}{3} \kjj \over 
\sqrt{(\koo - \frac 53 \kjj )^2 - 4 (\koo + \frac 13 \kjj )^2} } \right)
{ e^{-R/\la_-} \over R},
\label{Greensoln1}
\eea
where the $\la_\pm$ constants are defined by
\beq
{ 1 \over (\la_{\pm})^2 } = u \pm v,
\label{lambdas}
\eeq
and they act like two distinct length scales.
The other Green function $\cG_2$ has a similar solution but it is $\cG_1$ that is proportional to the Newtonian potential.

We note the contrast of this result \rf{Greensoln1} with previous results.
First, 
unlike the standard Yukawa potential, 
we have $2$ length scales.
Second, 
the amplitudes of the two terms vary depending on the values of the coefficients.
In particular, 
we find that these amplitudes could take on large values (i.e., much greater than unity) for a narrow range of coefficient ratios $\kjj / \koo$, even
if the coefficients themselves are ``small".
This is in contrast to standard assumptions of the smallness of Lorentz-violating effects.  
Note that the length scales would also be small, 
so such large Lorentz-breaking forces could escape detection in long-range tests. 

For practical evaluation 
over distributions of matter, 
such as those used in experiment, 
one would take the integral 
of the Green functions
over the smooth matter distributions 
$\rh(\vec r^\prime)$ as usual.
A more complete treatment of the these types 
of solutions is forthcoming.

\section*{Acknowledgments}
We acknowledge the support of the NSF, 
grant numbers 1806871 and 2207734, 
Embry-Riddle Aeronautical University, and the Arizona NASA Space Grant Consortium.

\end{document}